%% file: 0.main.tex
\documentclass[manuscript,screen]{acmart}

\AtBeginDocument{%
  \providecommand\BibTeX{{%
    \normalfont B\kern-0.5em{\scshape i\kern-0.25em b}\kern-0.8em\TeX}}}

\setcopyright{acmcopyright}
\copyrightyear{2018}
\acmYear{2018}
\acmDOI{10.1145/1122445.1122456}

\usepackage{booktabs} 
\usepackage{url}

\usepackage{caption}
\usepackage[all]{nowidow}
\usepackage{wrapfig}
\usepackage{array}
\usepackage{arydshln}
\usepackage{tabularx}
\usepackage{multirow}
\usepackage{arydshln}
\newcolumntype{L}[1]{>{\raggedright\let\newline\\\arraybackslash\hspace{0pt}}m{#1}}
\newcolumntype{C}[1]{>{\centering\let\newline\\\arraybackslash\hspace{0pt}}m{#1}}
\newcolumntype{R}[1]{>{\raggedleft\let\newline\\\arraybackslash\hspace{0pt}}m{#1}}

\usepackage{wrapfig,lipsum,booktabs} 

\def\authnotes{1}
\newcounter{notectr}[section]
\newcommand{\thenote}{\thesubsection.\arabic{notectr}\refstepcounter{notectr}}

\newcommand{\note}[2]{$\ll$#1~\thenote: #2$\gg$}
\newcommand{\cnote}[1]{\ifnum\authnotes=1 \textcolor{blue}{\note{Comment:}{#1}}\fi}

\copyrightyear{2026}
\acmYear{2026}
\setcopyright{acmlicensed}\acmConference[CHI '26]{CHI}{Nov 5-9, 2026}{USA}
\acmBooktitle{CHI'26, Nov, 2026, USA}
\acmPrice{15.00}
\acmDOI{10.1145/3491102.43}
\acmISBN{978-1-4503-9157-3/22/04}

\begin{document}




\title{Who Guides Whom? Negotiating Care, Religious Authority, and Technology in Bangladeshi Muslim Families}

\author{Susmita Biswas}
\affiliation{%
  \institution{United International University}
  \country{Bangladesh}}
\email{susmita.biswas.ador@gmail.com}

\author{Saiful Islam Tuhin}
\affiliation{%
  \institution{United International University}
  \country{Bangladesh}}
\email{ saiful.islam4@g.bracu.ac.bd}

\author{Faria Soroni}
\affiliation{%
  \institution{United International University}
  \country{Bangladesh}}
\email{soroni.faria@gmail.com}

\author{Sharifa Sultana}
\affiliation{%
  \institution{University of Illinois Urbana-Champaign}
  \country{USA}}
\email{sharifas@illinois.edu}

\author{Jannatun Noor}
\affiliation{%
  \institution{United International University}
  \country{Bangladesh}}
\email{jannatun@cse.uiu.ac.bd}

\renewcommand{\shortauthors}{Susmita. et al.}

\begin{abstract}

Digital technologies are reshaping how families negotiate care, authority, and responsibility, yet these dynamics remain understudied in religiously situated family life. Through interviews with 38 adult members of Bangladeshi Muslim families, we examine how family members guide one another across religious, digital, and AI-mediated contexts. We find that authority did not simply follow age or family hierarchy: older members often remained sources of religious and familial guidance, while younger members increasingly provided expertise around digital risks, information credibility, and AI. Families also actively shaped digital environments and negotiated the role AI could occupy in everyday relationships. We develop the concepts of household epistemic governance, counter-curation, and domain-dependent authority to explain these shifting relations of knowledge and responsibility. Our findings suggest opportunities for family technologies that support negotiated boundaries, credible information practices, and culturally situated forms of care.

\end{abstract}






\settopmatter{printfolios=true}

\maketitle

\input{1.intro}
\input{2.lit}

\input{3.methods}

\input{4.context}
\input{5.find}
\input{6.discussion}

\input{6.lim-fw-con}



\bibliographystyle{ACM-Reference-Format}
\bibliography{sample_base}

\end{document}

%% file: 1.intro.tex
\section{Introduction}
Digital technologies are increasingly intertwined with how families communicate, coordinate responsibilities, provide guidance, and navigate everyday concerns. Smartphones, social media, messaging applications, parental controls, location-sharing tools, recommendation systems, and, more recently, generative AI can shape how family members support one another, guide younger members, manage perceived risks, and make decisions around care and safety \cite{mols2023family,druga2022familyai,wen2025familiesvisionof,nichols2022digitally}. These technologies are taken up within relationships already shaped by trust, authority, privacy, responsibility, and expectations of care. Their use is therefore rarely an individual matter alone; it becomes part of the relational processes through which family members negotiate responsibilities, boundaries, dependence, and forms of support.

HCI scholarship has examined important dimensions of family technology use, including parental mediation, privacy and security, caregiving, and intergenerational differences in digital expertise \cite{livingstone2008parental,ghosh2018safety,tang2022never}. In South Asian and Muslim contexts, prior studies have explored more specific issues such as gendered access to technology, privacy practices, and religiously situated technology use \cite{sultana2018design,ahmed2017privacy,rifat2021purdah,ibtasam2021religious}. These studies provide valuable contextual insight, but they offer only partial views of how technology becomes entangled with the broader organization of family life. We still understand comparatively little about how multiple forms of guidance, responsibility, authority, and technology use are negotiated together within the same household.

Drawing on prior work on parental mediation, family surveillance, and joint family oversight \cite{clark2011parental,mols2023family,akter2022joint}, we use \textit{family governance} to describe the everyday processes through which family members establish, negotiate, and revise expectations around behaviour, relationships, responsibility, safety, and decision-making. We use the term broadly to include not only monitoring or rule enforcement, but also guidance, explanation, trust, care, and the gradual transfer of responsibility across generations and family roles. In Muslim households, these processes may also draw on religious commitments that shape ideas of care, upbringing, appropriate conduct, and responsibility, while still allowing different interpretations across people, generations, and circumstances \cite{white2012religion,dwinandita2024islamicchildparenting,douglass2022socialmediaand,rifat2021purdah}. This framing allows us to examine family technology use as part of a wider relational process, rather than treating it as a set of isolated decisions about devices or platforms.

Against this backdrop, Bangladesh offers a particularly compelling setting for examining family governance. Muslim identities and practices are deeply embedded in the country's social fabric, reflecting the religious affiliation of the large majority of its population \cite{white2012religion}, while everyday family life is also being reshaped by rapidly changing digital practices. Rather than treating these religious, social, and technological dimensions as separate influences, we examine how they become negotiated within family relationships themselves. Our focus is not on defining a singular model of the ``Muslim family,'' but on understanding how family members make sense of responsibility, guidance, boundaries, and technological change within their own relationships and circumstances.
To examine these dynamics, we ask,

\begin{itemize}
\item \textbf{RQ1:} How are family responsibilities, religious values, authority, and autonomy negotiated within Bangladeshi Muslim households?

\item \textbf{RQ2:} How do digital technologies and AI become involved in family practices of guidance, care, protection, trust, and decision-making?

\item \textbf{RQ3:} What culturally and religiously appropriate design approaches could better support negotiated and responsible technology use within these households?
\end{itemize}

Our findings show that family governance was negotiated rather than fixed, with authority shifting across relationships, gender, maturity, and domains of expertise even when family members shared religious commitments. Digital technologies and AI became active sites of this negotiation, as families curated information environments, redistributed digital expertise, and set limits on how socially or authoritatively AI should behave. These findings point toward family technologies that support revisable, reason-bearing boundaries and domain-scoped authority rather than imposing uniform control or positioning AI as a final source of judgment.


Our study contributes to HCI in three ways. First, we extend family-technology research beyond parent-child mediation by showing how care, authority, responsibility, and technological expertise are negotiated across multiple generations and family roles in Bangladeshi Muslim households. Second, we show how religious commitments, family relationships, and digital expertise become intertwined in shaping who is trusted, who guides whom, and how technological decisions are made within the household. Third, we show how AI can shape existing family relationships while families actively negotiate the relational role AI is permitted to occupy, including its authority, intimacy, and influence over judgment. Because participants frequently used religious and culturally specific terminology when describing family practices, we provide brief definitions of recurring terms in \autoref{tab:religious_terms}. These definitions reflect how the terms are used in this study rather than exhaustive theological meanings.
\begin{table}[!ht]
\centering
\caption{Religious and culturally situated terms used in this study}
\label{tab:religious_terms}
\Description{
A two-column table defining religious and culturally situated terms used in the study. 
Talim refers to a religious teaching or study gathering where participants learn about and discuss Islamic teachings. 
Zikr refers to the remembrance of Allah, often practiced through repeated prayers, Qur'anic recitation, or devotional phrases. 
Fajr is the obligatory dawn prayer and the first of the five daily Islamic prayers. 
Purdah refers to practices concerning modesty, privacy, gendered interaction, and, in some contexts, veiling or spatial separation. 
Amanah refers to trust, responsibility, or a moral obligation to safeguard what has been entrusted to a person. 
Gheebat means backbiting, particularly speaking negatively about another person in their absence. 
Ayat refers to a verse of the Qur'an. 
Surah refers to a chapter of the Qur'an. 
Qul means ``Say'' in Arabic; in this study, the participant phrase ``three Qul-type surahs'' refers to Surah Al-Ikhlas, Surah Al-Falaq, and Surah An-Nas. 
The Qur'an is the central scripture of Islam, believed by Muslims to contain the revelation of Allah conveyed to the Prophet Muhammad. 
Hujur is a commonly used Bangladeshi honorific for an Islamic religious teacher, preacher, or scholar. 
Moulobi is a South Asian term commonly used for a Muslim religious scholar or teacher. 
Sheikh is an Arabic honorific or title that, in this study, refers to a religious scholar or preacher. 
Mahram refers to a person whom one is permanently prohibited from marrying because of close kinship, marriage, or certain other recognized familial relationships. 
Imaan refers to faith or religious belief in Islam, including one's sense of commitment to and confidence in that faith. 
Alhamdulillah is an Arabic expression meaning ``praise be to Allah,'' commonly used to express gratitude, relief, or appreciation. 
Mosque refers to a Muslim place of worship used for communal prayer, religious learning, and other community activities. 
The table note clarifies that these definitions reflect how the terms are used in this paper and in participants' accounts, and are not intended as exhaustive theological or legal definitions.
}
\small
\renewcommand{\arraystretch}{1.08}
\begin{tabularx}{\textwidth}{@{}p{0.17\textwidth}X@{}}
\toprule
\textbf{Term} & \textbf{Meaning as used in this study} \\
\midrule

\textit{Talim} & 
A religious teaching or study gathering where participants learn about and discuss Islamic teachings. \\

\textit{Zikr} & 
The remembrance of Allah, often practiced through repeated prayers, Qur'anic recitation, or devotional phrases. \\

\textit{Fajr} & 
The obligatory dawn prayer and the first of the five daily Islamic prayers. \\

\textit{Purdah} & 
Practices concerning modesty, privacy, gendered interaction, and, in some contexts, veiling or spatial separation \cite{rifat2021purdah}. \\

\textit{Amanah} & 
Trust, responsibility, or a moral obligation to safeguard what has been entrusted to a person.  \cite{rifat2021purdah}. \\

\textit{Gheebat} & 
Backbiting, particularly speaking negatively about another person in their absence. \cite{rifat2021purdah}. \\

\textit{Ayat} & 
A verse of the Qur'an. \\

\textit{Surah} & 
A chapter of the Qur'an. \\

\textit{Qul} / ``Qul-type surahs'' & 
\textit{Qul} means ``Say'' in Arabic. In this study, the participant phrase ``three Qul-type surahs'' refers to Surah Al-Ikhlas, Surah Al-Falaq, and Surah An-Nas. \\

Qur'an & 
The central scripture of Islam, believed by Muslims to contain the revelation of Allah conveyed to the Prophet Muhammad (PBUH) \cite{mattson2008story}. \\

\textit{Hujur} & 
A commonly used Bangladeshi honorific for an Islamic religious teacher, preacher, or scholar. \\

\textit{Moulobi} & 
A South Asian term commonly used for a Muslim religious scholar or teacher. \\

\textit{Sheikh} & 
An Arabic honorific or title that, in this study, refers to a religious scholar or preacher. \\

\textit{Mahram} & 
A person whom one is permanently prohibited from marrying because of close kinship, marriage, or certain other recognized familial relationships. \\

\textit{Imaan} & 
Faith or religious belief in Islam, including one's sense of commitment to and confidence in that faith. \\

\textit{Alhamdulillah} & 
An Arabic expression meaning ``praise be to Allah,'' commonly used to express gratitude, relief, or appreciation. \\

\textit{Mosque} &
A Muslim place of worship used for communal prayer, religious learning, and other community activities. \\

\bottomrule
\end{tabularx}

\vspace{2pt}
\begin{minipage}{0.98\textwidth}
\footnotesize
\textit{Note: Definitions reflect how these terms are used in this paper and in participants' accounts. They are intended to support interpretation of the study rather than provide exhaustive theological or legal definitions. Meanings and practices associated with some terms may vary across communities, traditions, and contexts.}
\end{minipage}
\end{table}



%% file: 2.lit.tex
\section{Related Work}
\label{sec:related-work}

We approach family technology use through \emph{domestic governance}: the relational work through which families negotiate responsibility, care, authority, safety, privacy, and appropriate conduct. This framing shifts attention from individual use or parent-child control to the relationships and institutions that give technology meaning, including schools, religious communities, platforms, and AI systems. We situate it across four literatures: Islamic values and digitally mediated authority; family care, safety, privacy, and autonomy; intergenerational mediation and knowledge practices; and bioecological and neo-ecological approaches to family technology.

\subsection{Muslim Families, Religious Values, and Domestic Authority}

\subsubsection{Religion, Family Life, and Technology}

Research on religion and technology spans digital religious practice, family education, virtual religious institutions, Qur'anic learning, culturally situated games, and algorithmically mediated religious content \cite{wahid2024exploring,rahman2025islamic,salam2025virtual,dalimunthe2023challenges,sari2024learning,din2023ramadan,jahan2026should}. Across these settings, technology changes where religious learning occurs, which sources become accessible, and how religious knowledge is encountered. Within families, religious commitments are similarly cultivated through parenting, habituation, learning, and everyday routines, while adapting to changing digital environments \cite{dwinandita2024islamicchildparenting,fitriyah2023islamic,abubakar2023parenting,dasopang2022millennial,rouzi2020establishing,bukhari2024challenges}. Technology therefore becomes both a resource for religious learning and a setting in which families negotiate appropriate conduct, supervision, and responsibility.

Research from South Asia shows how these negotiations are shaped through family and social relations. Access may depend on relatives and shared devices, while gender, mobility, stigma, reputation, and unequal access shape participation and privacy \cite{ibtasam2019my,sultana2021opaque,sultana2018design,ahmed2017privacy,paudel2023deep}. Studies of Pakistani teenagers, immigrant families, and online communities similarly show privacy, security, and participation being negotiated alongside religion, gender, family expectations, and social visibility \cite{mustafa2023pakistaniteensand,naveed2026,usman2025}. Work on patriarchal social-media dynamics and women-only digital spaces further demonstrates how digital participation can require trust, moderation, and boundary management \cite{younas2020patriarchy,ammari2022moderation}. Together, this work positions technology use within family relations and social accountability rather than as an individual interaction with a device.

\subsubsection{Situated Religious Authority and Digital Participation}

HCI research further shows that religious commitments should be understood as situated forms of reasoning rather than background variables. Prior scholarship calls for greater engagement with religious beliefs, values, and epistemological concerns as part of lived technological practice \cite{ibtasam2021religious,noor2026decoding}, while work in Bangladesh shows how faith, moral worlds, and locally meaningful knowledge interact with computing and data-driven systems \cite{sultana2019witchcraft,sultana2020parareligious,sultana2021chasing}. Research with pious Muslim communities demonstrates how concepts such as \textit{purdah}, \textit{amanah}, and \textit{gheebat} inform privacy and surveillance \cite{rifat2021purdah}, while scholarship on Bangladeshi family life cautions against treating religious commitment and modern aspiration as opposites \cite{white2012religion}. These studies show that religious commitments become consequential through situated interpretation rather than producing one fixed pattern of behaviour.

Digital media further complicate religious authority by expanding the people, institutions, and platforms through which religious knowledge is encountered \cite{douglass2022socialmediaand,evolvi2017hybridmuslimidentities}. Recommendation systems can also shape exposure to religious content while raising concerns about relevance, authenticity, and misguidance \cite{jahan2026should}. Yet we know less about what happens when these questions enter family relationships themselves: how religious commitments are interpreted across generations, how competing sources of authority are negotiated, and how responsibility for access, guidance, and technological decision-making is distributed within the household.

\subsection{Negotiating Care, Safety, Privacy, and Relational Authority}

\subsubsection{Safety, Monitoring, and Negotiated Boundaries}

Parental-mediation research shows that families use a range of strategies, including discussion, restriction, co-use, and monitoring, and adapt these practices to children's age, digital skills, online experiences, and perceived risks \cite{livingstone2008parental,clark2011parental,nikken2013developingscalesto,symons2017qualitative,rodriguezdedios2018astudyof}. These practices often involve balancing opportunities for learning and participation online against concerns about digital risk and the need to guide children's technology use \cite{livingstone2017maximizing,cabellohutt2017onlineopportunitiesand,katz2017contextualizingconnectivit,sciacca2022parental}. Similarly, parents and young people may differ in how they understand appropriate involvement, monitoring, privacy, and self-regulation, even when they share concerns about online safety \cite{erickson2016boundaries,wisniewski2017middle,ghosh2018matter,ghosh2018safety}. These tensions have motivated HCI work that moves beyond restriction toward communication, cooperation, transparency, and negotiated oversight \cite{badillo2020beyond,wang2021protection,akter2022joint,stoilova2023doparentalcontrol}. Existing literature further frames protection alongside privacy, participation, information access, and children's developing capacities \cite{livingstone2014global,stoilova2021privacy,uncrc2021gc25}.

Protective intent, however, does not determine how monitoring is experienced. Location tracking and family surveillance can signify care, reassurance, coordination, control, or distrust depending on the relationship \cite{widmer2021ambiguities,mols2023family,zhangkennedy2016fromnosylittle}. Parental-control tools may also introduce privacy, security, and misuse risks \cite{maier2025surveillance}, while technology-facilitated abuse research shows how devices, accounts, and location data can enable coercive monitoring \cite{douglas2018technologyfacilitateddomes,brookfield2024technology,slupska2021threatmodelingintimate}. Family safety is therefore not simply a matter of increasing technological visibility, but of negotiating when and how protective authority is appropriate.

\subsubsection{Care Beyond Parent--Child Control}

Research in Bangladesh situates these negotiations within broader material and institutional conditions. Studies document privacy challenges around shared phones, adolescent internet risks, problematic internet use, and parental mediation \cite{ahmed2017privacy,ahmed2019everyone,chandrima2020adolescentproblematicinter,islam2020problematicinternetuse,wohab2015aconceptualframework}. Responses to harm may also involve families, communities, and institutions \cite{sultana2022shishushurokkha}, while work on child-friendly schools highlights institutional responsibilities for children's protection \cite{tusriyanto2022theimplementationof}. Safety practices are therefore distributed across relationships and settings rather than contained within the household.

Care also includes how adults themselves use technology. Research on technoference and family media use shows that caregivers' device practices can shape attention, stress, parenting interactions, and parent--child relationships, while cautioning against reducing these effects to screen time alone \cite{mcdaniel2018technoferencelongitudinala,barr2020beyond,modecki2020tuningintothe,komanchuk2023impactsofparental}. Together, they establish digital safety as a relational problem involving protection, privacy, autonomy, and reciprocal responsibility. Yet it remains predominantly centred on parents and minor children. We know less about how technological care and authority are negotiated among spouses, adult children and parents, siblings, and older relatives, particularly when these responsibilities also carry religious significance.

\subsection{Intergenerational Mediation, Digital Care, and Epistemic Work}

Technology use within families is often mediated through other people rather than accomplished by isolated users. HCI4D research describes \emph{intermediated use}, where a more knowledgeable person enables another's participation through translation, assistance, or surrogate action \cite{3sambasivan2010intermediated}. Studies of family Internet use and older adults' ICT engagement similarly show that digital participation depends on relatives, social relationships, prior experience, instruction, and factors such as digital anxiety \cite{eynon2014familydynamicsand,neves2013comingofold,cheng2021bridgingthedigital,ma2020bridgingthedigital,zhou2022thedigitaldivide,niu2026digital}. Family-centred work further connects such mediation to care, showing how digital communication sustains kinship, intergenerational relationships support digital literacy and future care, and younger relatives protect older adults from online fraud \cite{yu2023mobilizingcarewechat,bai2025digital,deng2025auntiepleasedont}.

Mediation also concerns what information should be trusted. Family members may assess unfamiliar sources, suspicious messages, algorithmically recommended content, or synthetic media before deciding whether to accept or share them. Research on Muslim digital life shows how online environments participate in religious identity formation \cite{evolvi2017hybridmuslimidentities}, while work on deepfakes and religious image manipulation highlights concerns around authenticity and misleading synthetic content \cite{sakinah2025islamic}. Digital mediation therefore involves not only helping someone access technology, but also helping them judge the credibility and significance of what they encounter.

Existing work further explains intermediated access, intergenerational learning, digital care, and online risk well, but says less about how these forms of mediation intersect when family members hold different kinds of expertise. In particular, we know less about how digital knowledge interacts with religious, moral, experiential, or familial authority, and how judgments about information are distributed across adult children and parents, spouses, siblings, and older relatives. This gap motivates our examination of domestic governance as more than one knowledgeable person simply assisting another.

\subsection{Family Governance Across Physical, Virtual, and AI-Mediated Ecologies}

\subsubsection{Bioecological and Neo-Ecological Perspectives}

Bronfenbrenner's bioecological theory explains development through the Process--Person--Context--Time (PPCT) model, emphasizing recurring interaction, individual differences, interconnected settings, broader cultural and institutional conditions, and change over time \cite{guy2020bronfenbrenner}. Neo-ecological theory extends this account to lives unfolding across physical and virtual microsystems, where interfaces, connectivity, platform processes, and digital literacy also shape experience \cite{navarro2023technologizing}. Ecological approaches have consequently been applied to digital family dynamics, technology-based interventions, parental technology use, digitally mediated play, and technological risk \cite{buarbuțua2023digital,folk2022anecodevelopmentalframewor,dworkin2022associationsamongparent,arnott2016ecological,gaitis2025extendedrealityenvironment}.

We use these perspectives as sensitizing rather than deterministic frameworks. They allow family technology use to be examined across relationships and settings rather than as an isolated technological effect, while keeping cultural values, institutions, platform environments, and changes in responsibility in view.

\subsubsection{Extending Family Ecologies to AI}

AI introduces new actors and interactions into these ecologies. Family-AI research examines parents and children learning about AI together, household uses of generative systems, and parental mediation of AI \cite{druga2022familyai,wen2025familiesvisionof,eira2025parentsperceptionsabout,zhang2025parentalmediationand}. Prior works conceptualize AI literacy through competencies such as recognizing AI, understanding its strengths and limitations, and critically evaluating issues such as misinformation \cite{long2020ai}. Child-rights and education scholarship and reports similarly emphasizes transparency, safeguards, literacy, participation, and equity \cite{kasneci2023chatgpt,unicef2025ai,unesco2023genai}. In religious contexts, algorithmically recommended and AI-generated material also raises questions of authenticity, authority, misguidance, and overreliance \cite{jahan2026should,jahan2026mind,sakinah2025islamic}.

Navarro further extends neo-ecological theory by treating AI heuristically as a ``person'' within PPCT analysis, allowing responsive systems and their designed characteristics to be examined as participants in repeated interaction \cite{navarro2026integrating}. Related work points to the effects AI may have on parenting, communication, socially responsive interaction, and technology-mediated support \cite{szondy2025artificialintelligenceai,alfeir2024dimensions,dray2025reviewofinnovative}. Yet less is known about how person--AI interactions subsequently enter existing family relationships and forms of authority, particularly where religious commitments and intergenerational responsibilities already shape how guidance is interpreted.

%% file: 3.methods.tex
\section{Methods}
\label{sec:methods}
 We conducted a qualitative study using semi-structured interviews with 38 adult members of Bangladeshi Muslim families to understand how religious values, care, authority, and technology use are negotiated within family life. We recruited participants across different generations and family roles, including those with experience raising children into adulthood, and analyzed the interview data using thematic analysis.




\subsection{Participant Recruitment and Interview Procedure}

We recruited 38 adult participants using purposive and snowball sampling to capture perspectives on religious values, care, and technology use within Bangladeshi Muslim family life. Participants were initially approached through the researchers' personal and community networks and through contacts connected with schools and faith-based educational institutions. Recruitment subsequently expanded through referrals from initial participants. Our purposive recruitment prioritized individuals with relevant experiences of family decision-making, guidance, caregiving, or technology use, while seeking variation in age, gender, occupation, marital status, geographical location, household arrangement, and family position where possible. This variation allowed us to examine family governance from different relational positions rather than from a single household role.

Participants were eligible if they were 18 years of age or older, identified with a Muslim family context, and were willing to discuss their everyday family relationships and technology practices. The resulting sample included participants across different generations and family roles and represented all eight administrative divisions of Bangladesh. ~\autoref{tab:participants} summarizes the participant characteristics.

Before each interview, participants were informed about the purpose of the study, participation procedures, voluntary nature of participation, intended use of the data, and recording practices. Participants then provided informed consent and completed a brief demographic questionnaire. We separately obtained permission to audio-record each interview before recording began. Participants were also informed that they could decline to answer any question or withdraw from the interview at any time.

We conducted semi-structured individual interviews online via in-person meetings or Google Meet. Data collection took place between May and August 2026. Each interview lasted approximately 40–60 minutes, yielding approximately 32–35 hours of recorded interview material across the 38 participants. The interviews explored family relationships and responsibilities, religious values and their role in everyday family life, care and authority, technology use and digital safety, inter-generational technological support and participants' experiences with emerging AI systems.

The semi-structured format provided consistency across these areas while allowing us to follow participants' experiences and probe issues that were particularly relevant to their family relationships and circumstances. Audio recordings were subsequently used for verbatim transcription and qualitative analysis in accordance with the study's data-handling procedures.

\begin{table*}[t]
  \caption{Demographic profile of the 38 study participants, showing distributions by gender, age, education, marital status, occupation, administrative division, and household arrangement.}
  \label{tab:participants}
  \Description{
    Table 2 summarizes the demographic characteristics of 38 participants. The sample included 20 women (52.6\%) and 18 men (47.4\%). Participants were predominantly young adults: 17 participants (44.7\%) were aged 18--24, 9 (23.7\%) were aged 25--30, 5 (13.2\%) were aged 31--40, 4 (10.5\%) were aged 41--50, and 3 (7.9\%) were aged 51 or above. In terms of education, 15 participants (39.5\%) had undergraduate-level education, 13 (34.2\%) had higher secondary education, 7 (18.4\%) had postgraduate education or above, 2 (5.3\%) had secondary education, and 1 (2.6\%) had primary education. Marital status included 21 single participants (55.3\%), 15 married participants (39.5\%), and 2 divorced participants (5.3\%). Occupations included 14 students (36.8\%), 12 participants in salaried employment (31.6\%), 5 homemakers (13.2\%), 4 unemployed participants (10.5\%), and one participant each in domestic work, retired government work, and daily-wage or informal work (2.6\% each). Participants came from all eight administrative divisions of Bangladesh: Chattogram 7 (18.4\%), Dhaka 6 (15.8\%), Khulna 5 (13.2\%), Rajshahi 5 (13.2\%), Barishal 4 (10.5\%), Rangpur 4 (10.5\%), Sylhet 4 (10.5\%), and Mymensingh 3 (7.9\%). Most participants lived in nuclear families, 29 (76.3\%), while 5 (13.2\%) lived in joint or extended families, 3 (7.9\%) lived with in-laws, and 1 (2.6\%) reported another household arrangement.
}
  \footnotesize
  \renewcommand{\arraystretch}{1.15}
  \setlength{\tabcolsep}{7pt}
  \centering
  \begin{minipage}[t]{0.49\textwidth}
    \centering
    \begin{tabular}{@{}l l r r@{}}
      \toprule
      \textbf{Characteristic} & \textbf{Distribution} & \textbf{$n$} & \textbf{\%} \\
      \midrule
      \textbf{Gender} & & & \\
      & Female & 20 & 52.6 \\
      & Male & 18 & 47.4 \\
      \midrule
      \textbf{Age} & & & \\
      & 18--24 & 17 & 44.7 \\
      & 25--30 & 9 & 23.7 \\
      & 31--40 & 5 & 13.2 \\
      & 41--50 & 4 & 10.5 \\
      & 51 and above & 3 & 7.9 \\
      \midrule
      \textbf{Education} & & & \\
      & Undergraduate & 15 & 39.5 \\
      & Higher Secondary & 13 & 34.2 \\
      & Postgraduate+ & 7 & 18.4 \\
      & Secondary & 2 & 5.3 \\
      & Primary & 1 & 2.6 \\
      \midrule
      \textbf{Marital status} & & & \\
      & Single & 21 & 55.3 \\
      & Married & 15 & 39.5 \\
      & Divorced & 2 & 5.3 \\
      \bottomrule
    \end{tabular}
  \end{minipage}
  \hfill
  \begin{minipage}[t]{0.49\textwidth}
    \centering
    \begin{tabular}{@{}l l r r@{}}
      \toprule
      \textbf{Characteristic} & \textbf{Distribution} & \textbf{$n$} & \textbf{\%} \\
      \midrule
      \textbf{Occupation} & & & \\
      & Student & 14 & 36.8 \\
      & Salaried employment & 12 & 31.6 \\
      & Homemaker & 5 & 13.2 \\
      & Unemployed & 4 & 10.5 \\
      & Domestic work & 1 & 2.6 \\
      & Retired govt.\ work & 1 & 2.6 \\
      & Daily-wage/informal & 1 & 2.6 \\
      \midrule
      \textbf{Division} & & & \\
      & Chattogram & 7 & 18.4 \\
      & Dhaka & 6 & 15.8 \\
      & Khulna & 5 & 13.2 \\
      & Rajshahi & 5 & 13.2 \\
      & Barishal & 4 & 10.5 \\
      & Rangpur & 4 & 10.5 \\
      & Sylhet & 4 & 10.5 \\
      & Mymensingh & 3 & 7.9 \\
      \midrule
      \textbf{Household arrangement} & & & \\
      & Nuclear family & 29 & 76.3 \\
      & Joint/extended family & 5 & 13.2 \\
      & Living with in-laws & 3 & 7.9 \\
      & Other arrangement & 1 & 2.6 \\
      \bottomrule
      
    \end{tabular}
  \end{minipage}
\end{table*}

\subsection{Data Analysis}

The authors independently transcribed the audio recordings and translated the Bengali transcripts into English. During transcription and translation, the authors revisited the recordings when necessary to preserve participants' intended meanings, expressions, and contextual nuances as closely as possible. Each participant was assigned a pseudonymous code before analysis.

We analyzed the interview data using thematic analysis \cite{clarke2017thematic}. The analysis began with repeated reading of the transcripts to familiarize ourselves with the data and generate initial codes grounded in participants' accounts. Initial codes included terms such as ``religious guidance through routine,'' ``explaining restrictions,'' ``gendered protection,'' ``trust replacing direct monitoring,'' ``keeping screens visible,'' ``filtering inappropriate content,'' ``checking religious credibility,'' ``younger members supporting older relatives online,'' and ``using AI for family advice.'' These codes were subsequently compared and grouped as broader patterns began to emerge across participants' accounts. These codes were able to capture a variety of experiences: family responsibilities, religious values, care and authority, trust and autonomy, digital practices, inter-generational support, safety concerns, and emerging uses of AI.

The coding and theme-development process was collaborative and iterative. The authors shared their analyses with each other to compare interpretations across transcripts, discuss emerging patterns, refine codes, and examine relationships among them. Overlapping codes were consolidated, distinctions between closely related concepts were clarified, and broader categories and candidate themes were developed. Throughout this process, we repeatedly returned to the transcripts to ensure that emerging interpretations remained grounded in participants' accounts.

The final themes were developed through multiple rounds of discussion and consensus. We attended both to recurring patterns and to differences across participants' relational positions, gender, age, household arrangements, and experiences with technology. The research questions informed the overall analytical focus, while the inductive process allowed themes and relationships not anticipated in advance to emerge from the data.

\subsection{Ethical Considerations and Positionality}

Our study protocol was reviewed and approved by the Institutional Review Board (IRB). All participants were adults and provided informed consent before the interview. They were informed about the purpose of the study, the voluntary nature of participation, and their right to decline any question or withdraw at any time. Given the sensitive nature of discussions around child safety, abuse, harassment, religious values, and family practices, interviewers avoided pressuring participants to disclose experiences they were uncomfortable discussing. The study focused on adult family members’ perspectives and did not involve interviews with children. Audio recordings and transcripts were stored securely, and personally identifiable information was removed prior to analysis and reporting.

The researchers involved in conducting the interviews were born and raised in Bangladesh and are familiar with the social and cultural contexts, family structures, and Muslim traditions within which participants described their parenting practices. This familiarity supported rapport with participants and helped us interpret culturally situated references and practices. At the same time, we do not assume that shared national or cultural backgrounds imply shared religious identities, beliefs, or levels of religious practice. We therefore remained attentive to how our individual backgrounds, experiences, and assumptions could shape both our interactions with participants and our interpretation of the data.

%% file: 4.context.tex

%% file: 5.find.tex
\section{Findings}
Our findings show that family governance was not organized through a single source of authority or a fixed set of rules. Instead, participants described an ongoing process in which religious values, care, trust, responsibility, and technology use were negotiated across generations and family relationships. Across participants’ accounts, this negotiation unfolded as a progression: families first cultivated values and expectations, then translated them into boundaries around everyday conduct - all the while relying on technology.

\subsection{Islamic Values and Intergenerational Family Guidance}
\subsubsection{Islamic Values in Everyday Family Life and Upbringing}

For many participants, family governance began long before an explicit rule was imposed: it began in the routines, stories, prayers, and everyday interactions through which children learned what the family considered right, P8\_female, for instance, described religion as providing recognizable boundaries, while emphasizing that adherence was developed through how these values were cultivated within the family rather than through direct coercion:

\begin{quote}
``Religion is also an important part of our family life, but it never stopped us from growing. No one forced us to follow these teachings; we chose to follow them ourselves. In my family, religion was taught more through gratitude.'' - P8\_female
\end{quote}

Rather than remembering religion primarily as a set of prohibitions, P8 described learning it through the emotional texture of family life, where gratitude and everyday practice gradually made religious expectations feel internal rather than externally imposed. 

Across interviews, participants described religious upbringing as shaping both everyday conduct and family routines. P2\_female described responsible parenting as ensuring that children \textit{``remain disciplined, that alongside their studies they also pray, and that they follow basic manners and etiquette.''} P5\_male similarly described religion as part of household routine, noting, \textit{`In the morning and evening, I try to play the three Qul-type surahs (see \autoref{tab:religious_terms}) at home,''} while other family members read the Qur'an (see \autoref{tab:religious_terms}), prayed, and practiced zikr (see \autoref{tab:religious_terms}). Together, these accounts show how religious practice became embedded in ordinary family life.

These routines were not limited to devotional practice; participants also drew on religious teachings to guide younger family members' everyday social conduct and manners. This became particularly visible as younger family members entered new stages of social life. Participants described drawing on religious teachings to guide everyday conduct around social interaction, attraction, and appropriate behaviour, while adapting such guidance to age and maturity. P32\_female, for example, recalled introducing this kind of guidance as her son approached adolescence:

\begin{quote}
``I did not discuss this with them when they were very young; I started talking about it when they were approaching their teenage years. I told them that they should lower their gaze. Allah has created girls beautifully, so naturally they may feel like looking at them, but if they continue looking, it would be sinful. Alhamdulillah (see \autoref{tab:religious_terms}), I think our surrounding environment also has an influence. Altogether, I feel they are receiving Islamic teachings in a good way.'' - P32\_female
\end{quote}

Her account shows how religious teachings became a practical guide for everyday conduct and self-regulation, introduced as relevant to younger family members' maturing social experiences.

As such guidance became more relevant to younger family members' changing experiences, participants also described adjusting how it was communicated. P24\_female and P10\_male noted a gradual shift from direct enforcement toward explanation, trust, and self-regulation as younger members matured. 

These findings show that \textbf{families adjusted both the focus and delivery of religious guidance as younger members' circumstances changed.} \textbf{With maturity, a greater responsibility was placed on their own understanding and judgment.}

\subsubsection{Intergenerational Continuity and Negotiated Interpretation}

Beyond guiding everyday conduct, sustaining religious family life also depended on how knowledge and values were carried across generations. Older relatives often passed moral and religious knowledge through repeated interaction and storytelling. P12\_female traced much of her own understanding to stories shared by her father, but questioned whether she could reproduce this with her granddaughter:

\begin{quote}
``I could not raise my granddaughter as I raised my daughters. They can't even write Bengali properly. How do I pass on my teachings, the knowledge that I received from my elders? I don't know! It pains me. My daughter is also frustrated that she cannot teach her daughter properly while living away from us. She has to go to work and leave her daughter.'' - P12\_female
\end{quote}

Her concern centered on the relationships through which religious knowledge had previously been passed on. Distance, work, language, and changing family arrangements reduced opportunities for relatives to model, explain, and reinforce these inherited values across generations.

Religious continuity also required families to judge which interpretations and sources could be trusted. P34\_female described Islam as providing a moral framework while expressing uncertainty about the competing religious voices encountered around it:

\begin{quote}
``I primarily see Islam as a source of moral and spiritual guidance rather than as a book of rigid rules. Islam gives me a framework of compassion, honesty, humility, and accountability. [...] What makes practicing Islam difficult is the rise of extreme voices and self-proclaimed religious speakers who distort the actual message. Everyone seems to have their own interpretation of the `correct' practice, so it can become genuinely confusing to distinguish authentic teachings from ideological noise.'' - P34\_female
\end{quote}

Her account highlights how preserving religious knowledge also involved evaluating the \textit{credibility} of competing sources and interpretations. This concern became especially visible when digital media entered family disagreements over religious practice.

P6\_male, for example, used an online religious video to support his interpretation during a disagreement with his mother about travelling for \textit{talim}:

\begin{quote}
``My mother goes to talim (see \autoref{tab:religious_terms}), and I did not agree with the way she was travelling for it. I even showed her a video of a moulobi (see \autoref{tab:religious_terms}) saying that it is not right to travel like this without a mahram (see \autoref{tab:religious_terms}). But even after watching it, my mother said that the hujur (see \autoref{tab:religious_terms}) was wrong. She still wanted to go to talim.'' - P6\_male
\end{quote}

Both positions remained within a religious frame, yet they differed over how religious guidance should apply in this particular situation. The online video introduced another interpretation into the disagreement, but did not resolve it; P6\_male's mother rejected the speaker's view and continued to rely on her own understanding of the practice.

Across these findings, religious authority within families was \textbf{neither singular nor fixed}: guidance shifted from enforcement toward trust as \textbf{younger family members matured}, and even when digital content entered generational disagreements, \textbf{whose interpretation prevailed} remained shaped by trust within the family rather than access to information alone.

\subsection{Calibrating Care, Authority, and Autonomy}

\subsubsection{Protection, Authority, and Gendered Boundaries}

Participants described family protection as involving judgments about risk and the amount of freedom, supervision, or restriction considered appropriate. These boundaries varied with age, gender, perceived vulnerability, and wider social expectations. Technology was one setting where such differences became particularly visible. P17\_female described a family in which concerns about romantic involvement led to stricter phone access for a daughter than for her siblings:

\begin{quote}
``I once told the parents of a child that all of their children, the older one, the younger one, were using mobile phones, so why was there so much of a problem only with the middle child, who was a girl? They were afraid that if the girl used a phone, she might fall in love and run away with someone. It later turned out that she actually was in a relationship, but that is not the point. The other children could also have fallen in love. The girl reacted strongly; she shouted, became angry, complained to us, and protested against the way she was being treated.'' - P17\_female
\end{quote}

The daughter's protest also shows that such boundaries were not always passively accepted. Younger family members could question restrictions they experienced as unequal, making protection itself a site of negotiation within the household.

A caregiver perspective made the rationale behind such gendered protection more explicit. P23\_female explained that although boys and girls should be raised with the same fundamental values, perceived differences in safety could require greater caution for girls:

\begin{quote}
``I don't think boys and girls should be raised differently in terms of basic values. But in Bangladesh, we do have to be a little more careful with girls because of the safety risks they face. It is not that they should be taught different values; the reality is that the risks can be different.'' - P23\_female
\end{quote}

Interestingly, participants did not necessarily attribute these differences directly to religion. P3\_female made this distinction explicit, describing restrictions around technology as arising ``more from family values than religion.'' Her account separated religious guidance from family expectations and wider social concerns that could also shape protective decisions.

Across these accounts, gendered boundaries emerged from perceived risk, family expectations, and social conditions, while family members could differ in how those boundaries were justified and experienced.

\subsubsection{Care as Calibrated: Protection, Experience, and Self-Regulation}

Participants also described care differently depending on their position within the family. Boundaries that caregivers understood through protection, communication, or responsibility could be remembered differently by those subject to them.

P12\_female, for example, remembered raising her daughters through openness and communication:

\begin{quote}
``My husband used to bring the girls from their schools sometimes, and they all shared information. He and I made family decisions through discussions. My girls were really free with their father; they even showed that they got proposed to by their boyfriends through video recordings to us. My girls were free to go anywhere as long as they told us where they were going.'' - P12\_female
\end{quote}

Her youngest daughter, P13\_female, remembered the same family environment through a different set of experiences:

\begin{quote}
``When I was in class 7, I was very good at sports. But they did not let me because my white skin would burn in the sun and I wouldn't get married. Obviously, the boys alongside me did not get treated this way. My parents did not let me go anywhere.'' - P13\_female
\end{quote}

The parent associated freedom with openness and mobility within agreed boundaries, while the daughter remembered opportunities those boundaries had restricted. Their accounts show how the same family boundary could be understood as protective by a caregiver and restrictive by a younger family member. 

Participants also described boundaries changing with maturity. P32\_female, for example, placed fewer restrictions on her 16-year-old son and increasingly explained the reasons behind limits, emphasizing that ``trust is very important.'' P10\_male described this shift more explicitly:

\begin{quote}
``If you put too many restrictions on children, they will eventually find some other way to do whatever they want. What is more important is teaching children the difference between right and wrong so that they can understand for themselves what they should and should not do. We did not really check their phones. [...] We did not need to track their locations either. Whenever they went somewhere, they told us where they were going, and we trusted them.'' - P10\_male
\end{quote}

As younger family members matured, some caregivers reduced direct supervision and placed greater emphasis on explanation, trust, and personal responsibility. Self-regulation therefore became one basis for deciding when greater responsibility could be transferred within the family.

Across this subsection, protective authority was \textbf{shaped by gender, family position, and perceived risk}, while \textbf{trust and self-regulation increasingly influenced how responsibility was shared} as younger family members matured. This negotiated pattern later became visible in how families approached digital, religious, and AI-mediated forms of guidance.

\subsection{Governing Digital Environments as Family Care}
Participants described actively shaping how digital technologies were accessed, used, and trusted within family life. These practices addressed safety and harmful content while also reflecting expectations around productive use, religious practice, and family responsibility. Digital care therefore involved \textbf{shaping technological environments} as well as monitoring individual behaviour.

\subsubsection{Structuring Digital Access, Visibility, and Use}

Participants described digital access as something to be managed rather than simply allowed or prohibited. Some explicitly questioned whether complete restriction was an effective form of protection. P3\_female, for example, argued that maintaining communication while reshaping its conditions could be preferable to cutting it off entirely:

\begin{quote}
``I think cutting off communication is an injustice. Stopping all contact is not a solution. Instead, the communication should be brought to a healthier level and managed differently, so that it is not misused.'' - P3\_female
\end{quote}

Her account frames protection as a matter of adjusting how access or communication takes place rather than removing it altogether.

At the same time, participants described stronger intervention when particular patterns of technology use became concerning. P3\_female further recalled that when her brother began using his phone until 3 a.m., her father \textit{``took the phone away, changed the password, and kept it aside,''} which she understood as being for his well-being. Here, restriction was introduced in response to a particular pattern of use rather than described as a permanent condition of access.

Other families structured technology use through visibility. P21\_male described allowing social media use while keeping it within shared household spaces:

\begin{quote}
``If under-18 children want to use social media, they can do it with permission, but it should be in an open space, like watching TV. They should not be allowed to close the door and use it, because then how will we know what they are watching?'' - P21\_male
\end{quote}

In this account, access remained available, but \textit{where} it occurred became part of how the family maintained oversight.

Religious practice could also shape digital boundaries through the organization of everyday family life. P22\_female described her sister's household, where screen visibility existed alongside religious learning and routines that structured activities throughout the day:

\begin{quote}
``My sister and her family follow an Islamic lifestyle. She is very clear about monitoring her children and positions the monitor or laptop so that the screen remains visible. She makes her children understand why such restrictions are needed. She also downloads lectures by different sheikhs (see \autoref{tab:religious_terms}) and recitations of surahs (see \autoref{tab:religious_terms}) online. Their entire lifestyle is Islamic, from waking up in the morning to going to the mosque (see \autoref{tab:religious_terms}). Because their daily routine is organized in this way, controlling the children's screen time happens quite naturally. They do not have to hide phones or other technologies from them or constantly keep devices away. The routine itself limits how much time they spend with technology.'' - P22\_female
\end{quote}

Here, religion shaped technology use through more than explicit rules about devices. Prayer, religious learning, shared activities, and expectations around everyday conduct organized their time and attention in ways that indirectly limited screen use. Religious practice therefore became part of the household environment through which digital boundaries were maintained.

Across these accounts, digital governance rarely meant blocking access outright. Families structured \textit{where}, \textit{when}, and \textit{under what conditions} technology could be used, embedding boundaries within broader household routines that made continuous restriction less necessary.

\subsubsection{Technical Safeguards and the Limits of Protection}

While previous findings showed how families organized technology use through access, shared space, and routine, participants also relied on technical safeguards to extend care beyond what they could directly observe. Filters, platform settings, and surveillance technologies offered additional forms of protection, but participants did not treat these systems as sufficient on their own. Families still had to decide whether safeguards were working as expected, what counted as harmful, and when technical protection required further intervention.

\begin{figure}[!ht]
    \centering

    \begin{minipage}[t]{0.8\textwidth}
        \centering
        \href{https://kahfguard.com/}{
            \includegraphics[width=\linewidth]{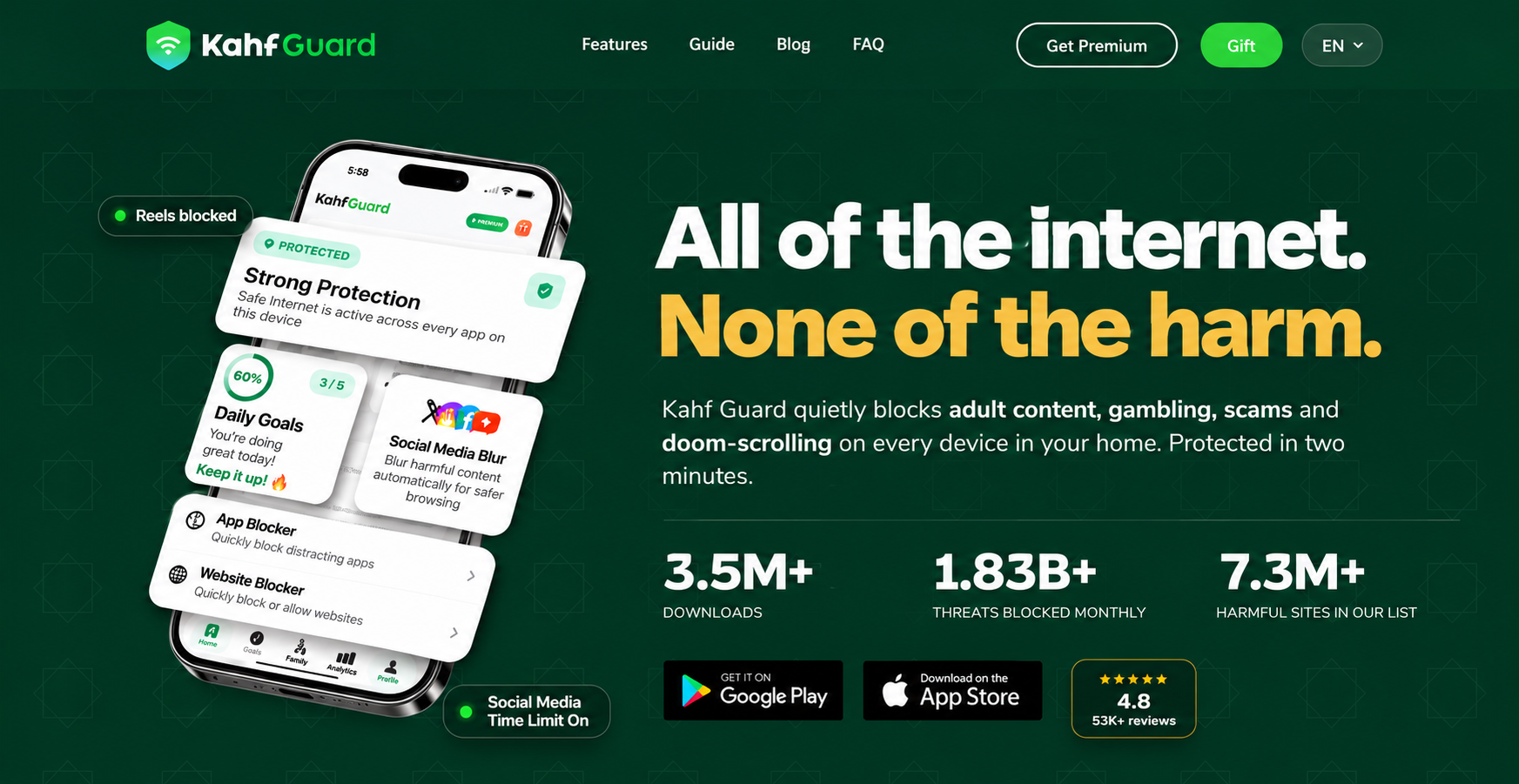}
        }
    \end{minipage}
    \hfill
    \begin{minipage}[t]{0.8\textwidth}
        \centering
        \href{https://kahfguard.com/}{
            \includegraphics[width=\linewidth]{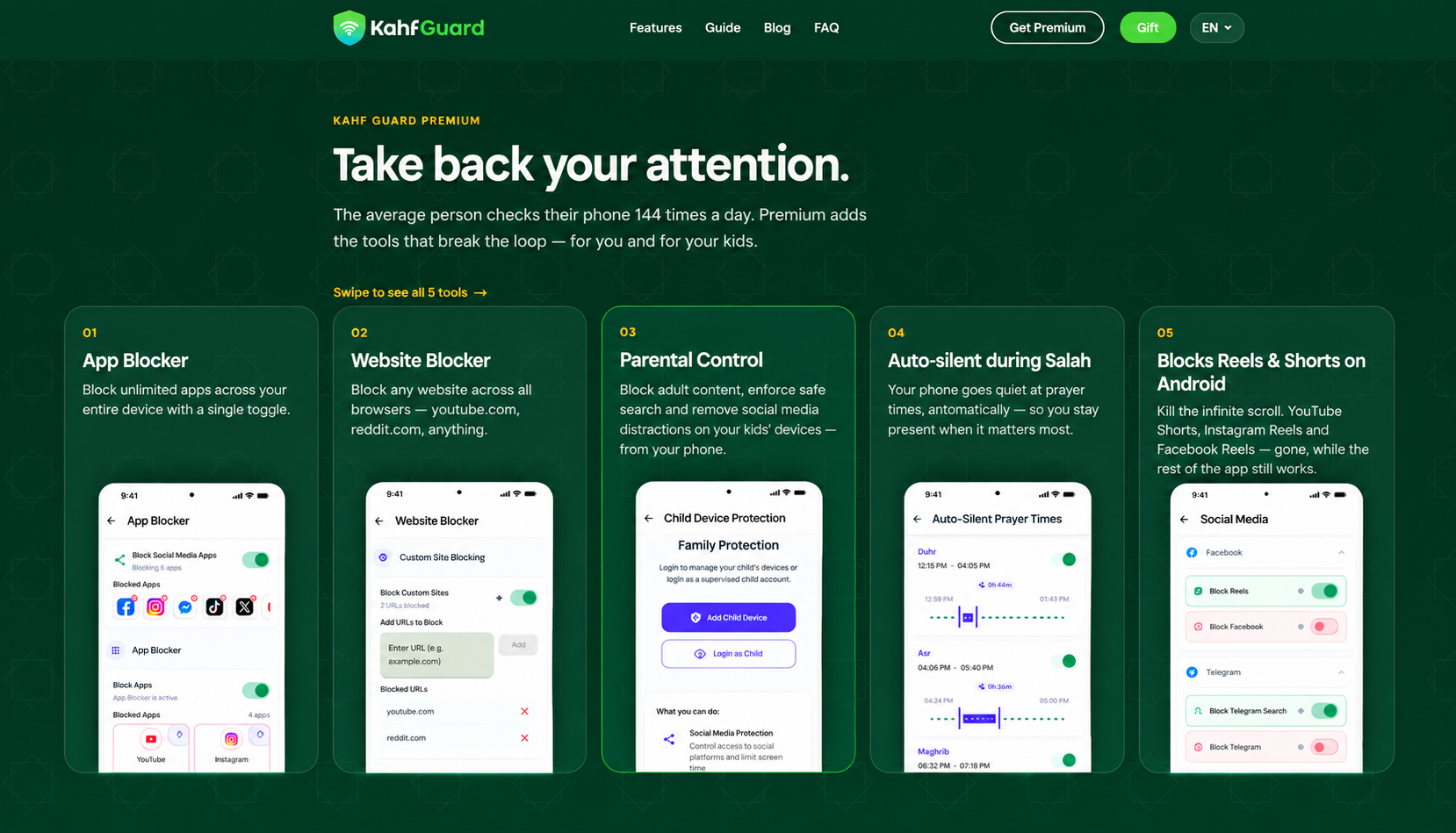}
        }
    \end{minipage}

    \caption{Kahf Guard as an example of a technical safeguard used to mediate 
religious and familial boundaries around digital content. The top image shows how 
the service presents protection from potentially harmful online content, while the 
bottom image presents configurable features including app and website blocking, 
parental controls, automatic silencing during Salah (prayer), and blocking of 
short-form social media content. Images from the \href{https://kahfguard.com/}{Kahf Guard website}; clicking the figure opens the website.}
\Description{
Figure 1 presents two screenshots from the Kahf Guard website, illustrating how the service describes its approach to filtering and managing online content. The upper screenshot introduces Kahf Guard as a tool for protection from harmful online content and shows a smartphone displaying the service. It describes protection from categories of content such as pornography or adult content, gambling, scams, and excessive or distracting online consumption. The lower screenshot presents several of Kahf Guard's configurable features. These include blocking selected applications, blocking websites, parental-control features, automatically silencing the device during Salah (prayer) times, and blocking short-form content such as Reels and Shorts on Android. Together, the screenshots show how Kahf Guard presents digital filtering and device-control features alongside religiously oriented practices and family-safety concerns. The figure is linked to the Kahf Guard website.
}
\label{fig:kahfguard}

\end{figure}

P30\_female, for example, used a filtering service to align online content with religious boundaries, but found that the system did not consistently reproduce the distinctions she expected:

\begin{quote}
``I downloaded Kahf Guard (See \autoref{fig:kahfguard}) to filter out content that I consider religiously inappropriate. But sometimes some content still gets through, and I am not sure why. I have tried adjusting and working with the app, but it does not seem completely reliable.'' - P30\_female
\end{quote}

Her account shows that technical protection required active caregiving work. P30\_female repeatedly adjusted the filter when it failed to reflect the religious boundaries she was trying to maintain, rather than treating the safeguard as a one-time solution. P15\_female similarly evaluated whether filtering aligned with the boundaries she expected. She described switching from her previous provider to Kahf Internet\footnote{Kahf Internet is a broadband service that applies router-level, DNS-based filtering across connected devices. According to its website, it blocks categories such as pornography, gambling, scam, malware, and phishing sites, and automatically enables safer search settings on services such as Google and YouTube. See: \url{https://kahfinternet.com/en/how-it-works/}.} and testing whether its filtering matched her expectations. She noted that it \textit{``blocks things like obscene content and gambling websites,''} while searches on pornography-related topics could still return research papers even when pornographic websites themselves were blocked. Together, these accounts show that families not only adopted technical safeguards, but also evaluated how well those systems enacted the boundaries they considered appropriate.

Some participants also located responsibility beyond individual households. P27\_male argued that platforms themselves should provide stronger protections against harmful and misleading material:

\begin{quote}
``I think we need community guidelines in both AI and Facebook that are more suitable for a country like Bangladesh [...] so that reports are taken seriously, and AI content guidelines can stop deepfakes. This may ease family members' minds as well.'' - P27\_male
\end{quote}

His account extends safeguarding responsibility to the platforms that shape what family members encounter, particularly when harmful or synthetic content may appear without being actively sought.

Technical safeguards also depended on how harm itself was understood. Participants did not always draw the same boundary between acceptable discipline and harmful treatment. P16\_male, for example, considered some physical discipline acceptable, stating that \textit{``sometimes some beating is necessary. But too much of anything is not good.''} P25\_male took a substantially different position, describing physical punishment in educational institutions as a safeguarding concern that should be formally prohibited:

\begin{quote}
``In various institutions, physical abuse, being beaten, is still a common experience. [...] Beating a student should clearly be treated as a punishable offence rather than something left to the discretion of individual institutions.'' - P25\_male
\end{quote}

The contrast is important because safeguarding was not only about making behaviour visible. Families and institutions also had to determine what practices should count as unacceptable harm and what forms of intervention should follow. Some participants therefore emphasized preparing younger family members to recognize uncomfortable situations and communicate them to trusted adults. P15\_female described beginning these conversations from an early age:

\begin{quote}
``When it comes to protecting them, I started talking to them about possible risks from a very young age. For example, child abuse can also happen to boys, so I discussed things like good touch and bad touch with them from childhood. Alhamdulillah, I have a very friendly relationship with my sons. They talk to me about their friends, what happened with them, and what their friends are doing. We discuss these things together.'' - P15\_female
\end{quote}

Her account places communication and disclosure alongside technical protection, allowing younger family members themselves to participate in recognizing and communicating possible harm.

These concerns became particularly significant when safeguarding extended beyond the household to institutions responsible for children's religious education. CCTV could provide families with additional visibility into such settings, but visibility did not necessarily produce trust. P30\_female described changing the institution her child attended after an earlier incident and remaining cautious even when the new institution provided CCTV:

\begin{quote}
``There was an incident at my eldest son's previous institution, and as soon as I found out, I changed his institution. The one they attend now has CCTV, but I still cannot fully trust them. [...] No matter how many CCTV cameras there are, I still do not feel comfortable leaving my children there overnight. The fear is always there. So every day, I bring them home at around 9 p.m. and take them back again in the morning.'' - P30\_female
\end{quote}

Her response shows the limits of surveillance as reassurance. CCTV increased what could potentially be observed, but it did not settle whether the institution itself could be trusted; P30\_female continued to rely on her own judgment and changed her caregiving routine accordingly.

Across these accounts, \textbf{technical safeguards extended families' capacity to protect, filter, and observe, but did not replace judgments about harm, trust, and intervention}. This interpretive work extended directly into how families evaluated the credibility of digital and religious content, discussed next.

\subsection{Digital Mediation of Religious and Intergenerational Authority}

Digital technologies changed how families encountered religious information and how expertise was distributed across generations. These shifts created new questions about credibility, guidance, and who was trusted to interpret particular forms of digital information.

\subsubsection{Religious Credibility and Household Curation}

 Participants often valued the greater access that digital media provided. P13\_female, for example, felt that technology had \textit{``broadened our religious knowledge''} by allowing people to check facts and meanings when they were uncertain. Greater access, however, also brought together scholars, edited clips, confident speakers, and algorithmically promoted content, making credibility harder to judge. P16\_male highlighted this problem in relation to younger users:

\begin{quote}
``A child may not know the difference between a qualified scholar and somebody who simply speaks confidently. Parents and teachers should recommend reliable sources rather than allowing likes and views to become the child's religious teacher. [...] They take videos of scholars, cut out important parts, and turn them into reels in ways that can completely change how the scholar appears.'' - P16\_male
\end{quote}

The concern was not simply exposure to “wrong” content. Digital platforms altered the cues through which religious authority was recognized, as popularity, confidence, short-form clips, and engagement metrics appeared alongside more established forms of expertise. Evaluating sources therefore became part of family religious guidance.

Some families responded by actively shaping what relatives encountered online. P15\_female, for instance, described intervening when her mother-in-law came across misleading Islamic clips:

\begin{quote}
``Another issue is that many Islamic clips are edited out of context, use misleading captions, or are created mainly to attract views. If I notice her watching something like that, I immediately explain the issue and give her an alternative source. I also subscribe her phone to authentic channels so that those videos appear more often. [...] I mark inappropriate or low-quality videos as `Not interested,' subscribe to better channels, and save useful videos for her. I have to repeat this every few months.'' - P15\_female
\end{quote}

Family members were not only recommending what to believe; they were also trying to influence what the platform would show next. Digital religious care thus involved a form of \textit{household counter-curation}, where families worked against misleading or engagement-driven content by steering feeds toward sources they considered more trustworthy.

\subsubsection{Intergenerational Digital Expertise and Reciprocal Authority}

 The work of evaluating and reshaping digital information also affected who became trusted to guide whom within the family. As digital expertise became more important, younger family members sometimes took on responsibilities that crossed conventional age-based authority. Older relatives often remained key sources of religious, moral, and familial knowledge, while depending on younger members to navigate unfamiliar technologies, suspicious messages, and online risks. Authority therefore did not simply shift across generations; instead, it became increasingly \textit{domain-specific}.

P19\_female described this form of intergenerational support as requiring attention to how assistance was offered:

\begin{quote}
``With my parents, the main protection is teaching them directly: do not click on a link if you are unsure of the source, and if something seems suspicious or unusual, speak to me about it before taking any action. My sibling and I have tried to create a safe environment where our parents can ask us for help with technology without feeling embarrassed.'' - P19\_female
\end{quote}

What mattered was also preserving the older family member's dignity. Participants described offering technical help in ways that avoided embarrassment and made it comfortable for older relatives to ask for assistance. P10\_male described a similar reversal when a caller claiming to represent his Hajj (The Islamic pilgrimage to Mecca) agency asked for financial information:

\begin{quote}
``Someone called me claiming to be from our Hajj agency and said we would receive a refund of 67,000 taka. They asked for my bank information, so I called my daughter. She realized it was a scam because they were asking for a number that should never be shared. Later, we contacted the actual agency and found out there was no such refund.'' - P10\_male
\end{quote}

In such moments, younger family members became digital interpreters for those who might otherwise hold greater authority within the family. Rather than displacing existing intergenerational roles, this \textbf{reconfigured authority as a domain-dependent process}: religious and familial guidance could continue to flow from older to younger generations, while \textbf{digital and financial expertise increasingly travelled in the opposite direction}. Intergenerational care thus became \textbf{a reciprocal distribution of expertise}, with different family members becoming guides, learners, and protectors depending on the problem at hand. These shifting patterns of trust and interpretation became more complex as AI itself began to enter family advice and decision-making.

\subsection{AI as an Emerging Relational Actor in Family Governance}

\subsubsection{AI as Advisor, Intermediary, and Governed Relationship}
Participants used AI for a range of everyday needs, from explaining unfamiliar information and checking documents to supporting study and navigating personal situations. In several cases, AI moved beyond information retrieval into moments of uncertainty, shaping how users interpreted and responded to emerging concerns. P18\_female illustrates this more relational use, where AI consultation extended into parenting and interpersonal decision-making,
\begin{quote}
``Whatever problem I have, I ask ChatGPT. For instance, if someone says something hurtful to me in a Facebook message, or gives me an unusual response, I ask AI what I should do. [...] Even when I am thinking about something myself, I wonder, `Is what I am thinking correct? Let me ask GPT and see.' [...] I also use AI for parenting. Once, my son's schoolteacher told me that he was very inattentive in class and often talked or mumbled to himself even while looking at the teacher. That worried me, so I asked GPT what I should do. GPT suggested that I spend more time with him, talk to him, and tell him stories. I followed those suggestions with my son for quite a few days, and now he does it much less than before.'' - P18\_female
\end{quote}
This positions AI as part of the interpretive work of family life. Beyond answering questions, it became a resource for framing concerns, considering possible responses, and informing decisions carried back into family relationships. Participants also expressed mixed feelings about this role. AI’s ready availability made it useful, while also raising questions about how much everyday judgment should be delegated to it.

Other participants responded to this possibility by deliberately limiting the kind of relationship AI could develop with family members. P15\_female did not prohibit AI, but attempted to configure how it interacted,

\begin{quote}
``Another concern I have is that people can become emotionally dependent on AI. I have set some restrictions for my children around this. Since AI can be customized, I configured it so that it does not simply agree with everything, but instead responds critically. I also prefer it not to behave too casually or too much like a friend; I want it to respond professionally. I often see screenshots where people share very emotional things with AI or talk to it as if they are talking to a close friend, and I do not like that.'' - P15\_female
\end{quote}

Here, governance moved beyond regulating access to AI toward shaping how AI was allowed to interact with family members. P15\_female's account defined boundaries around agreeableness, intimacy, disclosure, and emotional dependence, treating AI's social role as something that could itself be configured and limited.

\subsubsection{AI Literacy and Religious Authenticity}
AI also complicated how families judged the authenticity of religious information. Participants were particularly concerned that synthetic content could appear credible to relatives who had less familiarity with generative AI. P22\_female described discovering errors in religious material that appeared to have been AI-generated,

\begin{quote}
``Older people, like my mother-in-law, tend to believe AI-generated content without verifying it. If they see something about Islam, they may simply accept it. [...] Once we shared a picture containing an ayat (Qur'anic verse) in a group. Later, we noticed there were mistakes in the ayat, and that made us realize it might have been AI-generated. For things like this, we should use reliable websites and sources.'' - P22\_female
\end{quote}

Here, the concern extended beyond generic misinformation. When synthetic content appeared as Qur'anic or Islamic material, errors could enter the very knowledge families sought to preserve and pass on. AI trust therefore became tied not only to factual accuracy, but also to \textit{religious authenticity} and the reliability of the source through which religious knowledge was encountered.

Responding to this uncertainty also revealed differences in AI literacy across generations. Some younger family members described themselves as more cautious about synthetic content than older relatives who were less familiar with how convincingly AI-generated material could imitate credible information. P19\_female described this difference in her own family,

\begin{quote}
`` My parents are not very aware of what AI-generated content actually is. I am more skeptical because AI-generated content has become so convincing that it is now a genuine vehicle for spreading misinformation.'' - P19\_female
\end{quote}

AI literacy was not only about using AI, but also about judging whether digital content could be trusted. Differences in AI knowledge meant that some family members took greater responsibility for checking suspicious or synthetic content. This became particularly consequential when such material carried religious claims, making verification part of both digital care and the wider responsibility of protecting religious knowledge.


AI became involved in family life as both an \textbf{everyday advisor} and a source whose \textbf{relational role required boundaries}, particularly around dependence, intimacy, and critical use. Participants also emphasized \textbf{AI literacy, verification, and religious authenticity}, especially when synthetic content could be mistaken for trustworthy religious information.

%% file: 6.discussion.tex
\section{Discussion}


Our findings show that family, religion, and technology did not operate as separate influences on domestic governance, but became intertwined in how family members guided one another, assessed risk, negotiated boundaries, and shaped technological relationships. Rather than asking only how ecological systems influence family members, we examine how family members actively organize relations across household, religious, and digital settings.

\subsection{Family Governance as an Ecological Process}

Our findings show that technology-mediated family governance in Bangladeshi Muslim households is best understood as an ecological process rather than as a set of isolated rules or technology practices. We use Bronfenbrenner's Process-Person-Context-Time (PPCT) model and its neo-ecological extension to explain how family governance emerged through recurring interaction, differently positioned family members, connected physical and virtual settings, and change over time \cite{guy2020bronfenbrenner,navarro2023technologizing}. We use these concepts as an interpretive lens rather than as a formal test of PPCT.

\subsubsection{Process: Governance Through Repeated Interaction}

In bioecological theory, development is shaped through recurring reciprocal interactions between people and the persons, objects, and symbols in their immediate environments \cite{guy2020bronfenbrenner}. Our participants similarly described family governance as something produced through repeated explanation, guidance, discussion, supervision, disagreement, and trust rather than through one-time rules. Religious values were taught through everyday routines and conversations; technology boundaries were explained, challenged, and adjusted; and supervision could gradually give way to self-regulation. Family governance therefore emerged through continuing interaction in which both the meaning of a boundary and the response to it could change.

\subsubsection{Person: Position Shapes Authority and Experience}

The PPCT model emphasizes that people enter interactions with different characteristics that shape how those interactions develop \cite{navarro2023technologizing}. In our findings, age, gender, family role, experience, religious knowledge, and digital expertise affected both what authority a family member could exercise and how a decision was experienced. The same restriction could be understood by a caregiver as protection and by a younger family member as lost opportunity. Likewise, an older relative could provide religious or familial guidance while depending on a younger relative for help with scams, digital content, or AI. Authority was therefore not attached uniformly to age or family position, but varied with the person, relationship, and issue involved.

\subsubsection{Context: Governance Across Physical and Virtual Settings}

Bioecological theory situates people within interconnected settings, while neo-ecological theory extends these relations to virtual microsystems \cite{guy2020bronfenbrenner,navarro2023technologizing}. Our findings show family governance moving across these settings. A concern raised at school could become a conversation or rule at home; information encountered on social media could change how a family judged a religious source or institution; and platform recommendations, filters, or AI responses could enter subsequent family decisions. Religious commitments also became meaningful through connections among household, educational, religious, and digital settings. Family governance therefore involved not only what happened within the home, but how families interpreted and acted on experiences arriving from other parts of their ecology.

\subsubsection{Time: Responsibility and Authority Change}

Time in the PPCT model captures continuity and change across relationships and the life course \cite{navarro2023technologizing}. Participants similarly described authority as revisable: restrictions could lessen with age, monitoring could give way to trust, and responsibility could shift toward self-regulation. We saw \textit{domain drift}, where trust in one's expertise in one area could gradually expand their authority into related areas. A younger relative first relied on for scam detection, for example, could later become a trusted source for broader financial or technological decisions. Although our retrospective data do not trace these shifts longitudinally, they suggest that family authority changes over time both within and across domains.

\subsection{Religion as an Enacted Ecology}

Bioecological and neo-ecological theory place broader belief systems and cultural values within the macrosystem \cite{guy2020bronfenbrenner,navarro2023technologizing}. Our findings complicate a purely macrosystemic reading of religion where participants enacted religious commitments within everyday interactions and technological decisions, making religion part of the interactional material of family life rather than only its cultural backdrop. We call this \textit{enacted ecology}.

\subsubsection{From Religious Commitment to Situated Practice}

Religious commitments became practical through everyday decisions about conduct, relationships, digital access, schooling, and family routines. Prior HCI work in Bangladesh similarly shows how faith and locally meaningful forms of reasoning become intertwined with technological practice \cite{sultana2019witchcraft,sultana2020parareligious,sultana2021chasing}; our findings locate this process within family ecologies, where religious commitments are continually interpreted through relationships, technologies, and changing circumstances rather than simply transmitted from macro-level values into behaviour.

\subsubsection{Shared Religion, Different Interpretations}

This enactment did not produce one uniform household response. Family members could share religious commitments while differing over trusted sources, appropriate restrictions, gendered expectations, or how a principle should apply in a particular situation. Prior work similarly shows that religious concepts can shape privacy and surveillance through locally situated interpretation \cite{rifat2021purdah}, while scholarship on Bangladeshi family life cautions against treating religious commitment and modern aspiration as opposites \cite{white2012religion}. Religion therefore provided a shared moral vocabulary without determining a single form of authority or behaviour.

This distinction matters for HCI because treating religion only as a broad cultural variable can obscure how it becomes consequential in everyday technological practice. Our findings suggest studying religious commitments across ecological levels, as broader orientations are enacted through family relationships and reshaped through digital and institutional settings. Similar dynamics may arise across other religious traditions, motivating comparative HCI research on how belief systems become differently enacted within everyday family ecologies.

\subsection{Household Epistemic Governance}

Greater access to digital information did not remove uncertainty from family life; it shifted part of family governance toward deciding what could be trusted. Neo-ecological theory treats virtual microsystems as environments jointly shaped by users, technologies, and platform processes \cite{navarro2023technologizing}. Our findings show families responding not only to information, but also to the systems that selected and amplified it.

\subsubsection{From Access to Credibility}

Participants questioned edited clips, unfamiliar speakers, engagement metrics, suspicious messages, and synthetic content, often comparing them against sources they considered more reliable. This was particularly consequential for religious information, where visibility or confidence did not necessarily signal authority. Prior work similarly shows that digital media can reshape religious authority and that recommendation systems affect how religious content is encountered and judged \cite{campbell2020contextualizing,jahan2026should}. We call this family work of checking, questioning, and deciding what should be trusted or shared \textit{household epistemic governance}.

\subsubsection{Authority Depends on the Knowledge at Stake}

Neo-ecological theory already recognizes that younger people's digital expertise can unsettle age-based hierarchies \cite{navarro2023technologizing}, while prior HCI work shows younger relatives supporting older adults' technology use \cite{tang2022never,martinez2022warm}. Our findings show that this did not simply reverse family authority. An older relative could remain trusted for religious, moral, or experiential guidance while depending on a younger relative for scams, unfamiliar content, or AI. We describe this as \textit{domain-dependent authority}: authority followed the knowledge relevant to the problem rather than extending automatically across the relationship.

\subsubsection{Counter-Curation as Relational Maintenance}

Families also reshaped what platforms presented by marking unwanted material, subscribing to trusted sources, and introducing alternatives. We call this \textit{counter-curation}. This work was relational as well as epistemic: participants corrected older relatives while avoiding embarrassment or direct confrontation, echoing prior work on tactful intergenerational digital assistance \cite{martinez2022warm,deng2025auntiepleasedont}.
Counter-curation was also recurring. Participants repeatedly adjusted feeds and revisited questionable sources as content changed, making it a form of relational maintenance that sustained both the information environment and the relationships through which correction remained possible \cite{horst2021digital}.

\subsection{AI as a Relational Intermediary}

Our findings show that AI could enter family governance without being used directly by every family member. Navarro's neo-ecological extension treats AI heuristically as a ``person'' within PPCT analysis because responsive systems can participate in repeated interactions and shape human activity through their designed characteristics \cite{navarro2026integrating}. We extend this proposition by showing how person--AI interaction can subsequently shape another human relationship.

\subsubsection{AI Literacy as a Family Concern}

Participants' accounts aligned with several dimensions of AI literacy identified by Long and Magerko, including recognizing AI-generated content, understanding its limitations, and critically evaluating misinformation \cite{long2020ai}. In our study, these competencies were situated within the relational and religious dynamics of Muslim households, where judgments about AI-generated information were intertwined with responsibilities for preserving credible religious knowledge. These competencies were also unevenly distributed across family members and became especially consequential when synthetic content carried religious claims. AI literacy therefore emerged not only as an individual capacity, but as part of how families collectively judged what information could be trusted.

\subsubsection{From Person-AI Interaction to Family Interaction}

Participants sometimes brought family concerns to AI, interpreted its response, and then acted on that interpretation within the original relationship. A concern about a child, for example, could move from parent to AI and then back into subsequent parent-child interaction. AI therefore did not need to interact directly with both family members to affect their relationship. It could operate as a \textit{relational intermediary}, shaping how one person understood and responded to another.

This significantly extends the neo-ecological account. The relevant process is not only repeated interaction between a person and an AI system, but also the movement of AI-mediated interpretation into an existing family relationship. AI can therefore connect interactions that occur in different relational settings, with consequences that extend beyond the immediate user.




\subsubsection{Disciplining AI into Relational Propriety}

Navarro's ``AI as person'' heuristic helps make AI's relational role visible \cite{navarro2026integrating}, but our findings complicate it as families did not simply accommodate AI as another relational actor; they also disciplined the kind of actor it was allowed to become. One participant explicitly configured AI to challenge rather than simply agree, preferred a professional over friend-like tone, and placed limits around emotional dependence. AI's acceptable role was therefore itself subject to norms around intimacy, disclosure, reliance, and judgment.

This matters because socially persuasive and highly agreeable AI can shape users' judgments and relationships \cite{zhang2025dark,cheng2026sycophantic}. Our findings suggest that families may respond by defining \textit{relational propriety} for AI, not only what it may do, but how close, agreeable, or authoritative it should be allowed to become. AI thus enters family governance both as an intermediary in human relationships and as a relational actor whose own position must be actively bounded.

\subsection{Designing for Accountable Relational Authority}

Taken together, our findings suggest that family technologies should not treat authority as a simple choice between individual autonomy and household control. Authority in our study was conditional; it depended on the relationship, the concern being addressed, the knowledge involved, and how responsibilities changed over time. We use \textit{accountable relational authority} to describe a design orientation in which technological authority is scoped, explainable, and open to revision.

\subsubsection{Make Boundaries Explainable and Revisable}

Family-safety technologies often encode authority through permissions, restrictions, monitoring, or access controls. Prior HCI research has shown the limitations of designs that privilege unilateral parental control over communication, autonomy, and joint oversight \cite{wisniewski2017middle,badillo2020beyond,akter2022joint}. Our findings suggest that a boundary should also communicate \textit{why} it exists, \textit{what} it applies to, and \textit{when} it should be reconsidered. We call these \textit{reason-bearing boundaries}. For example, a restriction could be connected to a concern about safety, privacy, religious commitment, or responsible use without requiring the system itself to decide what that value should mean. This allows technology to support family reasoning without turning culturally or religiously situated judgments into fixed platform rules.

\subsubsection{Keep Authority Scoped to the Relevant Domain}

Family technologies should also distinguish between different forms of authority. Someone trusted to verify a suspicious link does not necessarily need access to another person's messages, while a family member providing religious or moral guidance need not control unrelated aspects of digital use. Systems could therefore separate roles such as \textit{advising}, \textit{assisting}, \textit{verifying}, and \textit{deciding}, with different permissions attached to each. This follows from our finding that expertise and authority were domain-dependent, and helps prevent legitimate responsibility in one area from silently expanding into broader surveillance or control.

\subsubsection{Return AI-Mediated Concerns to Human Deliberation}

Conversational AI introduces a different design problem because its responses can enter relationships that the system does not directly observe. When users seek advice about family concerns, AI should therefore avoid presenting itself as the final adjudicator. Instead, it could expose uncertainty, clarify assumptions, identify relevant sources, and offer alternative interpretations before returning the issue to the people involved. Such support can also strengthen AI literacy by helping users recognize AI-generated content, understand its limitations, and critically evaluate misinformation \cite{long2020ai}.  We describe this as a \textit{relational handoff}. This is particularly important for religiously consequential questions, where AI may support access to information without claiming religious authority \cite{jahan2026mind}. User-configurable boundaries around disclosure, intimacy, and highly affirming responses may further help people define the relational role they want AI to occupy \cite{zhang2025dark,cheng2026sycophantic}.

These directions shift family technology design from \textit{encoding values} toward supporting their \textit{articulation and negotiation}. Rather than defining one model of family life, systems can make reasons, responsibilities, limits, and changing authority visible while leaving interpretation to the people involved. This approach may also apply beyond religious households wherever care, expertise, and authority are distributed across family members.

%% file: 6.lim-fw-con.tex
\section{Reflections and Conclusion}

\subsection{Limitations and Future Work}

Our study has some limitations. Retrospective interviews limit our ability to examine how family practices change over time; longitudinal work could better trace shifts in trust, supervision, religious guidance, and technology use. In addition, our findings primarily reflect adult perspectives, suggesting the need for multi-member household studies. While grounded in Bangladeshi Muslim households, the study surfaces relational patterns and design considerations that may be transferable to other settings when adapted to their cultural, socioeconomic, and religious conditions. Future work can further examine how these design directions operate across different family contexts, particularly in supporting care while remaining attentive to surveillance, autonomy, and unequal authority.


\subsection{Conclusion}

This study examined how technology becomes part of everyday family governance in Bangladeshi Muslim households. We show that care, authority, trust, religious values, and technology use are negotiated across generations, relationships, and digital settings rather than governed through fixed rules. Drawing on bioecological and neo-ecological perspectives, we highlight household epistemic governance, counter-curation, domain-dependent authority, and AI's emerging role as a relational intermediary through which families navigate information, expertise, and responsibility. For HCI, these findings point toward family technologies that support negotiation, revisable responsibility, credible information, and culturally situated forms of care rather than relying primarily on control or surveillance.

